\documentclass[aps,pra,10pt,twocolumn,floatfix]{revtex4-1}
\usepackage{amsmath,amssymb,bm,graphicx}
\usepackage{times}

\usepackage{hyperref}
\hypersetup{
colorlinks=true,
linkcolor=blue,          
citecolor=blue,          
filecolor=magenta,       
urlcolor=cyan
}

\begin{document}

\title{Tuning the relaxation dynamics of ultracold atoms in a lattice with an optical cavity}
\author{E.~I.~Rodr\'iguez Chiacchio and A.~Nunnenkamp}
\affiliation{Cavendish Laboratory, University of Cambridge, Cambridge CB3 0HE, United Kingdom}

\date{\today}

\begin{abstract}
We investigate the out-of-equilibrium dynamics of ultracold atoms trapped in an optical lattice and loaded into an optical resonator that is driven transversely. We derive an effective quantum master equation for weak atom-light coupling that can be brought into Lindblad form both in the bad and good cavity limits. In the so-called bad cavity regime, we find that the steady state is always that of infinite temperature, but that the relaxation dynamics can be highly non-trivial. For small hopping, the interplay between dissipation and strong interactions generally leads to anomalous diffusion in the space of atomic configurations. However, for a fine-tuned ratio of cavity-mediated and on-site interactions, we discover a limit featuring normal diffusion. In contrast, for large hopping and vanishing on-site interactions, the system can be described by a linear rate equation leading to an exponential approach of the infinite-temperature steady state. Finally, in the good cavity regime, we show that for vanishing on-site interactions, the system allows for optical pumping between momentum mode pairs enabling cavity cooling.
\end{abstract}

\maketitle

\section{Introduction}

Ultracold atomic gases have proven to be an ideal playground for simulating quantum many-body physics \cite{RevBloch,RevBloch2012}. The Bose-Hubbard (BH) model \cite{JakschPRL1998} is a paradigmatic example for which the superfluid-Mott insulator phase transition has been observed experimentally \cite{Bloch}.

More recently, ultracold atomic gases have been loaded inside optical cavities \cite{Review}. These set-ups are very interesting experimentally as cavity photon losses provide a window for in-situ monitoring of the system. The global light-matter coupling mediates long-range interactions among the atoms. This has led to the experimental realization \cite{Esslinger, EsslingerDickeSB, EsslingerDickeRoton, EsslingerDickeFluct, HemmerichPRL2014, HemmerichPNAS2014} of the well-known Dicke phase transition \cite{Dicke, Domokos, Bhaseen1, Bhaseen2} with the observation of coherent emission of the cavity field and atomic self-organization. Additionally, the intrinsic dissipative nature of cavities makes these set-ups an interesting arena for the study of systems out of equilibrium \cite{Ciuti5, Ciuti, Ciuti3}. This is exciting as it has been shown that such systems can exhibit novel universality classes and dynamical critical behavior \cite{Diehl2, Diehl}.

The description of ultracold atoms in optical lattices coupled to a cavity mode can be based on a BH-type model whose physics is determined by the competition between kinetic energy, on-site, and infinite-range interactions. The ground-state phase diagram of this system  and quench dynamics were explored experimentally \cite{HemmerichPRLdec2015, Esslinger2,EsslingerMet}. Further theoretical studies have since analyzed the excitation spectrum of the system \cite{Huber}, considered the effects of incommensurate lattices \cite{Morigi}, and treated the trapping potential explicitly \cite{Sundar}. This has provided a detailed understanding of the coherent phenomena of the system. However, little is known about its far-from-equilibrium nature. Driven-dissipative systems can exhibit notably different phase diagrams from their equilibrium counterparts \cite{Ciuti5} and the interplay of interactions and dissipation can lead to unusual relaxation dynamics \cite{Cai,KollathPRL2012, KollathPRL2013, KollathPRL2015}.

\begin{figure}[t]
\centering
\includegraphics[width=\columnwidth]{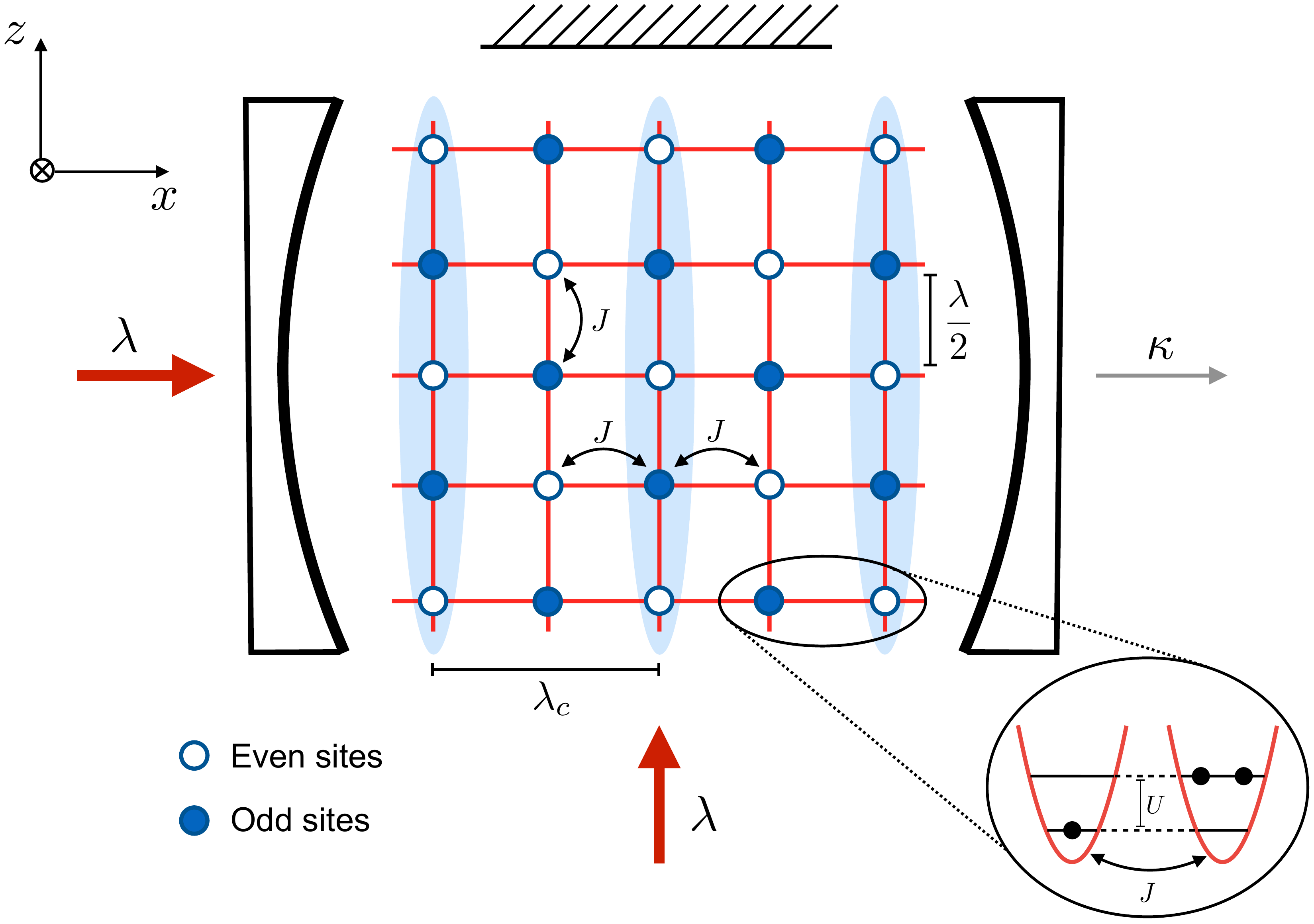}
\caption{\label{FigModel} Schematic representation of the system. A gas of ultracold atoms is placed inside a high-finesse cavity with cavity wavelength $\lambda_{c}$ and photon loss rate $\kappa$. Two lasers of wavelength $\lambda$ (red arrows) form a periodic lattice potential of wavelength $\frac{\lambda}{2}$. Atoms can hop across the lattice in the $x$-$z$ plane with an amplitude $J$ and are subject to an on-site interaction $U$. The laser in the transverse direction also acts as an external pump which can scatter photons off the atoms into the cavity and vice versa.}
\end{figure}

In this paper we investigate the effects of dissipation in the set-up realized in Ref.~\cite{Esslinger2}. The model is introduced in Sec.~\ref{Mod}. For weak atom-light coupling, we derive an effective quantum master equation for the atoms by adiabatically eliminating the cavity field (Sec.~\ref{SecAdiab}). In the bad cavity limit (Sec.~\ref{BCR}), where the cavity field adiabatically follows the atoms, we recover the effective Hamiltonian studied in Refs.~\cite{Esslinger2, Huber, Morigi, Sundar}, but supplemented with measurement-induced dephasing between states of different imbalance between the even and odd sites. The steady state is that of infinite temperature, however, the relaxation dynamics can change drastically: For small hopping (Sec.~\ref{BCRsJ}), the atoms explore configuration space in anomalous diffusion \cite{KollathPRL2012, KollathPRL2013, KollathPRL2015}. In contrast, if short- and long-range interactions are of the same order, their effects can cancel giving rise to normal diffusion. For large hopping and vanishing short-ranged interactions (Sec.~\ref{BCRlJ}), we show that the dynamics can be reduced to a linear rate equation predicting an exponential relaxation to the infinite-temperature state. Finally, in Sec.~\ref{GCR}, we study the good cavity limit for zero on-site interactions, where optical pumping between momentum pairs enables cavity-assisted cooling, similar to the one observed in Ref.~\cite{HemmerichNat2012}. We conclude in Sec.~\ref{Conc}.

\section{Model}
\label{Mod}

We consider a gas of ultracold atoms placed inside an optical cavity, in the presence of an external optical lattice and driven by a pump laser in the direction transverse to the cavity axis (see Fig.~\ref{FigModel}). We focus on the case where the cavity wavelength $\lambda_{c}$ is commensurate with that of the lasers forming the lattice $\lambda_{c}=\lambda$. Such a system can be described with the BH model including an atom-light interaction term \cite{Esslinger2}, which in the rotating frame of the pump reads ($\hbar=1$)
\begin{equation}
\label{eq:er001}
\hat{H} = \hat{H}_{U} + \hat{H}_{J} + \hat{H}_{\Delta} + g(\hat{a}^{\dagger}+\hat{a})\hat{\Phi},
\end{equation}
with $\hat{H}_{U} + \hat{H}_{J}= \frac{U}{2} \sum_{i} \hat{n}_{i} \left( \hat{n}_{i} - 1 \right) -  J \sum_{\langle i,j \rangle} \hat{b}^{\dagger}_{i}\hat{b}_{j}$ the BH Hamiltonian, being $\hat{n}_{i}=\hat{b}^{\dagger}_{i} \hat{b}_{i}$ the atomic number operator at site $i$ and $\hat{\Phi} =\sum_{e}\hat{n}_{e}-\sum_{o}\hat{n}_{o}$, where $e$/$o$ stands for even/odd sites. The operators $\hat a$ and $\hat b$ obey bosonic commutation relations. Here, $J$ is the nearest neighbor hopping amplitude and $U$ the on-site interaction strength. The third term represents the cavity photons $\hat{H}_{\Delta}= -\Delta \hat{a}^{\dagger} \hat{a}$, where $\Delta=\omega_{\textrm{pump}}-\omega_{\textrm{cav}}$ is the laser detuning with respect to the cavity mode, and the last term is the light-matter coupling induced by the pump. This interaction corresponds to photon scattering between the pump field and the cavity mode, which is determined by the atomic distribution across the lattice.

We include cavity losses using the quantum master equation
\begin{equation}
\label{eq:er1b} \partial_{t}\hat{\rho}=\mathcal{L}\hat{\rho} \equiv (-i)[\hat{H},\hat{\rho}] + \kappa \mathcal{D}[\hat{a}](\hat{\rho}) 
\end{equation}
where $\mathcal{D}[\hat{a}](\star) \equiv \hat{a} \ \star \ \hat{a}^{\dagger} - \frac{1}{2} \ \{ \hat{a}^{\dagger}\hat{a},  \star \}$ is the standard dissipator and $\kappa$ is the rate of photon loss. As we are most interested in characterizing the atomic dynamics, we focus on the case of weak light-matter coupling $g$. In this regime, we can adiabatically eliminate the photonic degrees of freedom and obtain an effective description for the atoms.

\section{Adiabatic elimination}
\label{SecAdiab}

Adiabatic elimination for equations of the form \eqref{eq:er1b} can be carried out using the Nakajima-Zwanzig formalism \cite{Gardiner}. This yields an effective equation of motion for the reduced atomic density matrix to second order in the light-matter coupling $g$
\begin{align}
\label{eq:er2}
\mathcal{P}\hat{\rho}_{\textrm{tot}}(t) = & \mathcal{P} \mathcal{L}_{\textrm{at}} \mathcal{P}\hat{\rho}_{\textrm{tot}}(t) \\
&+ g^2 \mathcal{P} \mathcal{L}_{\textrm{int}} \int_{0}^{\infty} dt' \ e^{(\mathcal{L}_{\textrm{at}}+\mathcal{L}_{\textrm{ph}})t'}  \mathcal{L}_{\textrm{int}} \mathcal{P}\hat{\rho}_{\textrm{tot}}(t-t') \nonumber,
\end{align}
where the projector $\mathcal{P}$ is defined as $\mathcal{P} \hat{\rho}_{\textrm{tot}}(t)= \textrm{Tr}_{\textrm{ph}} [\hat{\rho}(t)] \otimes \hat{\rho}^{\textrm{ss}}_{\textrm{ph}}$, with $\hat{\rho}^{\textrm{ss}}_{\textrm{ph}}$ the steady state density matrix for the photons in the absence of coupling. The Liouvillian terms are defined as $\mathcal{L}_{\textrm{at}} \hat{\rho}_{\textrm{tot}}=(-i)[\hat{H}_{U}+\hat{H}_{J},\hat{\rho}_{\textrm{tot}}]$, $\mathcal{L}_{\textrm{ph}} \hat{\rho}_{\textrm{tot}}=(-i)[\hat{H}_{\Delta},\hat{\rho}_{\textrm{tot}}] + \kappa \mathcal{D}[\hat{a}](\hat{\rho}_{\textrm{tot}})$ and $\mathcal{L}_{\textrm{int}} \hat{\rho}_{\textrm{tot}}=(-i)[\hat{H}_{\textrm{int}},\hat{\rho}_{\textrm{tot}}]$.

To take the trace over the photon sector, we need to consider correlation functions of the form $\langle \hat{\xi}(t)  \hat{\xi}(t') \rangle$, with $\hat{\xi}=\hat{a}^{\dagger}+\hat{a}$ and where $\langle \dots \rangle$ denotes an average for $g=0$. These can be obtained by considering the bath coupled to the light field as a zero-temperature source of white noise, i.e.~the only non-vanishing correlation function is $\langle \hat{a}_{\textrm{in}}(t) \hat{a}^{\dagger}_{\textrm{in}}(t') \rangle = \delta(t-t')$, where $\hat{a}_{\textrm{in}}$ and $\hat{a}^{\dagger}_{\textrm{in}}$ are input noise operators \cite{Girvin}. This yields $\langle \hat{\xi}(t)  \hat{\xi}(t') \rangle = \langle \hat{a}(t)  \hat{a}^{\dagger}(t') \rangle = e^{-\frac{\kappa}{2} |t-t'| + i \Delta (t-t')}$. Inserting this into \eqref{eq:er2} and tracing over the cavity mode, we obtain
\begin{align}
\label{eq:er3}
\dot{\hat{\rho}}_{\textrm{at}}(t) = &\mathcal{L}_{\textrm{at}} \hat{\rho}_{\textrm{at}}(t) \nonumber \\
&- g^{2} \int^{\infty}_{0}  dt' \ \Big( e^{-\frac{\kappa}{2} |t'| + i \Delta t'}  [\hat{\Phi}(0),\hat{\Phi}(-t')\hat{\rho}_{\textrm{at}}(t)] \nonumber \\
&\quad  + e^{-\frac{\kappa}{2} |t'| - i \Delta t'} [\hat{\rho}_{\textrm{at}}(t)\hat{\Phi}(-t'),\hat{\Phi}(0)]    \Big) ,
\end{align}
where $\hat{\rho}_{\textrm{at}}(t)=\textrm{Tr}_{\textrm{ph}} [\hat{\rho}(t)]$, $\hat{\Phi}(t)=e^{i\hat{H}_{\textrm{at}}t} \hat{\Phi} e^{-i\hat{H}_{\textrm{at}}t}$ and we have made use of the Markov approximation. Since \eqref{eq:er3} is an integro-differential equation, it is difficult to use in practice.

The imbalance operator $\hat{\Phi}=\sum_{e}\hat{n}_{e}-\sum_{o}\hat{n}_{o}$ commutes with the on-site interactions $[\hat{H}_{U},\hat{\Phi}]=0$, so its free evolution is $\hat{\Phi}(t)=e^{i\hat{H}_{J}t}\hat{\Phi}e^{-i\hat{H}_{J}t}$. In the quasimomentum basis $\hat{b}_{k}= \frac{1}{\sqrt{K}} \sum_{j} \hat{b}_{j} e^{ikj}$, where $j$ denotes the lattice sites, $K$ is the total number of sites, and $k=\frac{2\pi}{K}n$ the quasimomentum with $n=0,\dots,K$ an integer, $\hat{H}_{J}=\sum_{k} \varepsilon_{k} \hat{b}_{k}^{\dagger} \hat{b}_{k}$ and $\hat{\Phi}=\sum_{k} \hat{b}^{\dagger}_{k} \hat{b}_{k-k_{\pi}}$, with $k_{\pi}=\pi$. We then have $\hat{\Phi}(t) = \sum_{k} e^{i(\varepsilon_{k}-\varepsilon_{k-k_{\pi}})t} \ \hat{b}^{\dagger}_{k}\hat{b}_{k-k_{\pi}}$. Using this property, we can integrate over $t'$ and obtain
\begin{align}
\label{eq:er4}
\dot{\hat{\rho}}_{\textrm{at}} =& (-i)[\hat{H}_{U} + \hat{H}_{J},\hat{\rho}_{\textrm{at}}] \nonumber \\
&+ g^{2} \sum_{k \in \textrm{BZ}} \Big\{G(-\varepsilon_{k}+\varepsilon_{k-k_{\pi}}) [\hat{b}^{\dagger}_{k} \hat{b}_{k-{k}_{\pi}} \hat{\rho}_{\textrm{at}},\hat{\Phi}] \nonumber \\
&\qquad \qquad - G^{*}(\varepsilon_{k}-\varepsilon_{k-k_{\pi}}) [\hat{\rho}_{\textrm{at}} \hat{b}^{\dagger}_{k} \hat{b}_{k-{k}_{\pi}},\hat{\Phi}] \Big\}
\end{align}
with
\begin{align}
\label{eq:er4b}
G(\omega)=&\int_{0}^{\infty} dt' \ e^{-\frac{\kappa}{2}|t'|+i(\Delta + \omega)t'} \nonumber \\
=&\frac{\kappa/2}{(\Delta + \omega)^2 + \kappa^{2}/4} + i\frac{\Delta + \omega}{(\Delta + \omega)^2 + \kappa^{2}/4} ,
\end{align}
where the sums over $k$ run over the first Brillouin zone (BZ) $k \in [-\pi,\pi)$. Equation \eqref{eq:er4} is a Markovian quantum master equation that is local in time. However, its non-Lindblad form makes its physical interpretation not straightforward. 

It is possible to bring \eqref{eq:er4} into Lindblad form in two different regimes: the bad cavity limit $J \ll (\kappa,|\Delta|)$, where $G(\omega)$ becomes independent of $\omega$, and in the good cavity limit, where terms of the form $(\hat{b}^{\dagger}_{k}\hat{b}_{k-k_{\pi}})^{2}$ can be safely neglected using a rotating wave approximation (RWA). In the following, we focus on the bad cavity limit and will discuss the good cavity limit in Sec.~\ref{GCR}.

\section{Bad cavity regime $J \ll (\kappa,|\Delta|)$}
\label{BCR}

In the bad cavity limit $J \ll (\kappa,|\Delta|)$, the cavity follows the dynamics of the atoms adiabatically, and $G(\omega)=\frac{\kappa/2}{\Delta^2 + \kappa^{2}/4} + i\frac{\Delta}{\Delta^2 + \kappa^{2}/4}$, which allows us to bring \eqref{eq:er4} into Lindblad form
\begin{equation}
\label{eq:er8}
\dot{\hat{\rho}}_{\textrm{at}} = (-i)[\hat{H}_{\textrm{eff}}, \hat{\rho}_{\textrm{at}}] + \gamma \mathcal{D}[\hat{\Phi}] ( \hat{\rho}_{\textrm{at}})
\end{equation}
with
\begin{equation}
\label{eq:er8b}
\hat{H}_{\textrm{eff}}=\hat{H}_{U}+ \hat{H}_{J}+\hat{H}_{U_{l}} ,
\end{equation}
where 
\begin{equation}
\label{eq:er8c}
\hat{H}_{U_{l}}=-\frac{U_{l}}{K} \left( \sum_{e}\hat{n}_{e}-\sum_{o}\hat{n}_{o} \right)^2 ,
\end{equation}
with
\begin{align}
\label{eq:er8d}
U_{l}&=-K g^{2} \textrm{Im}[G(\varepsilon_{k}-\varepsilon_{k-k_{\pi}})]=-K\frac{g^2\Delta}{\Delta^{2}+\frac{\kappa^{2}}{4} }\\
\gamma&= 2g^{2} \textrm{Re}[G(\varepsilon_{k}-\varepsilon_{k-k_{\pi}})]=\frac{g^2\kappa}{\Delta^{2}+\frac{\kappa^{2}}{4}} .
\end{align}
The effective Hamiltonian (\ref{eq:er8b}) features hopping, on-site interactions, and infinite-range interactions $\hat{H}_{U_{l}}$, a consequence of the global coupling of all atoms to the single-mode cavity field. The Hamiltonian (\ref{eq:er8b}) has been previously studied in Refs.~\cite{Huber, Morigi, Sundar} where it was shown that the ground-state phase diagram exhibits four phases classified by the presence or absence of atomic coherence and even-odd imbalance.

Crucially, we find that there is dephasing between atomic configurations corresponding to different imbalance with rate $\gamma$, which comes as the dissipative counterpart to the coherent long-range interactions $\hat H_{U_l}$. Since $\hat{\Phi}$ is Hermitian, the dissipator can be rearranged as a commutator with the density matrix $\mathcal{D}[\hat{\Phi}]\hat{\rho}_{\textrm{at}}=\frac{1}{2}[\hat{\Phi},[\hat{\rho}_{\textrm{at}},\hat{\Phi} ] ]$, meaning that the steady state $\hat{\rho}^{\textrm{ss}}_{\textrm{at}}$ needs to obey $[\hat{\rho}^{\textrm{ss}}_{\textrm{at}},\hat{H}_{\textrm{eff}}]=[\hat{\rho}^{\textrm{ss}}_{\textrm{at}},\hat{\Phi}]=0$.

For $J=0$, the Hamiltonian and the relaxation operator satisfy $[\hat{H}_{\textrm{eff}},\hat{\Phi}]=0$, making the quantum master equation \eqref{eq:er8} exactly solvable. Since both $\hat{H}_{\textrm{eff}}$ and $\hat{\Phi}$ are diagonal in the site basis, the steady states correspond to number states in this basis. The dissipator in \eqref{eq:er8} eliminates coherences between states associated with different eigenvalues of $\hat{\Phi}$, but does not affect coherences between states where these are equal. For a general initial state, expressed in number state basis, this means the density matrix can be decomposed in blocks, corresponding to different eigenvalues of $\hat{\Phi}$, which remain unchanged by dynamical evolution and will preserve coherence in the long time limit. Steady states with non-vanishing coherences are usually referred to as decoherence free subspaces \cite{Lidar} and have been subject of much investigation due to  potential applications for quantum computing \cite{Lidar2}.

For $J\not=0$, $\hat{H}_{\textrm{eff}}$ and $\hat{\Phi}$ do not commute anymore $[\hat{H}_{\textrm{eff}},\hat{\Phi}] \neq 0$. From this, it follows that the steady state is unique and $\hat{\rho}^{\textrm{ss}}_{\textrm{at}} \propto \openone$. This corresponds to a steady state of the form $\hat{\rho}(t=\infty)=\frac{1}{M}\sum_{\mathbf{n}} | \mathbf{n} \rangle \langle \mathbf{n}|$, where $\mathbf{n}=(n_{1},n_{2},...,n_{K})$ denotes a specific atomic configuration and $M=\binom{K+N-1}{N}$, i.e.~the external pump eventually heats the system up to the completely mixed (infinite-temperature) steady state.

In the following, we will study the relaxation dynamics towards the infinite-temperature steady state. We find two different regimes: for small hopping the interplay between interactions and dissipation leads to (normal and anomalous) diffusion \cite{KollathPRL2012, KollathPRL2013}, while for large hopping the steady state is approached exponentially.

\subsection{Small hopping limit  $J \ll (U,U_{l},\gamma)$}
\label{BCRsJ}

Following \cite{KollathPRL2012, KollathPRL2013, KollathPRL2015}, we start our analysis by perturbatively eliminating the density matrix coherences, given that for strong interactions and in the presence of dephasing these should not play an important role in the evolution of the system. This yields an effective description in terms of the diagonal elements of $\hat{\rho}_{\textrm{at}}$. We then simplify the problem by introducing a mean-field decomposition and obtain analytical results in the limit of large particle filling, where we can derive a continuum description for the equations of motion.

The equations of motion for the coherences can be approximated to 
\begin{align}
\label{eq:er09a} \partial_{t} \rho_{\mathbf{n}}^{\mathbf{n}+e^{1}_{i,j}} &\simeq \left[ -iu(n_{i}-n_{j}+1) -2\gamma \right]  \rho_{\mathbf{n}}^{\mathbf{n}+e^{1}_{i,j}} \\
&+ iJ \sqrt{n_{l}+1} \sqrt{n_{r}}\left( \rho_{\mathbf{n}}^{\mathbf{n}} - \rho_{\mathbf{n}+e^{1}_{i,j}}^{\mathbf{n}+e^{1}_{i,j}}\right), \nonumber 
\end{align}
where $e^{d}_{i,j}$ is a vector whose $i$th component is equal to $d$, its $j$th component equal to $-d$, and the rest are equal to 0. We have also introduced the parameter $u=U-4U_{l}/K$, which effectively parameterizes the difference in strength between the two types of interaction. The approximate sign stands for having ignored the coupling to other coherences, which barely influences the dynamics in this limit. We focus on this set of coherences since they are the only ones coupled to the diagonal elements. Using $J \ll (U,U_{l},\gamma)$ we can integrate \eqref{eq:er09a} to obtain

\begin{equation}
\label{eq:er09b} \rho_{\mathbf{n}}^{\mathbf{n}+e^{1}_{i,j}} \simeq \frac{J\sqrt{n_{l}+1}\sqrt{n_{r}}}{u(n_{i}-n_{j}+d)-2i\gamma} \left( \rho_{\mathbf{n}}^{\mathbf{n}} - \rho_{\mathbf{n}+e^{1}_{i,j}}^{\mathbf{n}+e^{1}_{i,j}}\right) ,
\end{equation}
where we have neglected the transient terms and kept terms up first order in  $(J/u,J/\gamma)$ (see Supplemental Material in \cite{KollathPRL2015} for further details). Plugging this in the equations of motion for the diagonal elements, we obtain

\begin{align}
\label{eq:er9}
\partial_{t} \rho_{\mathbf{n}}^{\mathbf{n}}  =  &4 \gamma J^{2} \sum_{\substack{\langle i,j \rangle \\ d= \pm1}} \frac{(n_{i}+\delta_{d,1})(n_{j}+ \delta_{d,-1})}{u^2(n_{i}-n_{j}+d)^{2} + 4 \gamma^{2}}\times \nonumber \\
&  \qquad \qquad  \qquad \qquad \left( \rho_{\mathbf{n}+e^{d}_{i,j}}^{\mathbf{n}+e^{d}_{i,j}} -\rho_{\mathbf{n}}^{\mathbf{n}} \right).
\end{align}

Note that short- and long-ranged interactions only influence the dynamics through the factor of $u$ in \eqref{eq:er9}. This indicates that even when the system is strongly interacting, these interactions can counteract each other, rendering their effects negligible and leaving $J$ and $\gamma$ as the only energy scales. We thus distinguish two different regimes, $uN/\gamma \gg 1$ and $uN/\gamma \ll 1$.

We simplify the form of \eqref{eq:er9} using a Gutzwiller ansatz 
\begin{equation}
\label{eq:er11}
\hat{\rho}(t) =  { \displaystyle \bigotimes_{j=1}^{K} } \left[ \sum_{n_{j}} \rho_{j}(n_{j},t) |n_{j} \rangle \langle n_{j}| \right] ,
\end{equation}
where we only make a distinction between even and odd sites $\rho_{e}(m,t) \neq \rho_{o}(m,t)$, but consider all even/odd to be equivalent among themselves $\rho_{e(o)}(m,t) =  \rho_{e'(o')}(m,t)$. Plugging this ansatz into \eqref{eq:er9} we obtain
\begin{align}
\label{eq:er12} &\partial_{t} \rho_{e(o)} (n,t) = 4zJ^2\gamma \sum_{\substack{m \\ d= \pm 1}} \frac{(n+\delta_{d,1})(m+ \delta_{d,-1})}{u^2(n-m+d)^{2} + 4 \gamma^{2}} \times  \nonumber \\ 
& \left[ \rho_{e(o)}(n+d,t)\rho_{o(e)}(m-d,t) - \rho_{e(o)}(n,t)\rho_{o(e)}(m,t) \right],
\end{align}
where $z$ is the coordination number and $f=N/K$ is the lattice filling. By integrating this equation numerically we can access all the properties of the system. In this language, the steady state of the system now adopts the form $\rho_{e(o)}(n,t=\infty) =  \frac{1}{M} \binom{K+N-n-2}{N-n}$. Considering the limit of large system size $K \rightarrow \infty$, this can be recasted as $\rho_{e(o)}(n,t=\infty) \simeq f^{n+1}/[f(1+f)^{n+1}]$ using Stirling's formula.

To explore \eqref{eq:er12} analytically, we follow Refs.~\cite{KollathPRL2012, KollathPRL2013, KollathPRL2015} by considering the limit of large filling $f$ and introducing a continuous variable $x=n/f$. The probability distributions are redefined as $p_{e(o)}(x=n/f,t)=f\rho_{e(o)}(n,t)$, with $p_{e(o)}((n+1)/f,t)= p_{e(o)}(x,t) + \partial_{x}[p_{e(o)}(x,t)]dx$ and the steady state given by $p_{e(o)}(x,\infty)=e^{-x}$.

We choose to focus first on the case where $uN/\gamma \gg 1$. In the continuum limit, the equations of motion \eqref{eq:er12} read
\begin{align}
\label{eq:er14}
\partial_{\tau_{1}} p_{e(o)}(x,\tau_{1}) = \partial_{x} \Big[& D_{o(e)}(x,\tau_{1}) \partial_{x}p_{e(o)}(x,\tau_{1}) \nonumber \\
&- F_{o(e)}(x,\tau_{1}) p_{e(o)}(x,\tau_{1}) \Big]
\end{align}
with
\begin{align}
\centering
\label{eq:er15}
D_{i}(x,\tau_{1}) &= \int_{0}^{\infty} \frac{xy p_{i}(y,\tau_{1})}{(x-y)^2 + \frac{4\gamma^2}{u^2f^2}} dy \\
F_{i}(x,\tau_{1}) &= \int_{0}^{\infty} \frac{xy \partial_{y}p_{i}(y,\tau_{1})}{(x-y)^2 + \frac{4\gamma^2}{u^2f^2}} dy \ ,
\end{align}
and where we have introduced a dimensionless time $\tau_{1}=t/t_{1}^{*}$, with  $t_{1}^{*} = \frac{u^2 f^2}{4zJ^2\gamma}$. Next, we analyze \eqref{eq:er14} in the limits where analytical solutions can be obtained.
 
For short times $\tau_{1} \ll 1$, we can ignore the effects of $F_{e(o)}(x,\tau_{1})$ by considering sharply peaked and symmetrically distributed initial conditions $p_{e(o)}(x,0)=\delta(x-x_{e(o)})$. The dynamics around these initial points $x_{e(o)}$ can then be approximated by
\begin{equation}
\label{eq:er15b}
\partial_{\tau_{1}} p_{e(o)}(x,\tau_{1}) =  \partial_{x} \left[ \left( \frac{x_{e} x_{o}}{(x-x_{o(e)})^2 + \frac{4\gamma^2}{u^2 f^2}} \right) \partial_{x}p_{e(o)}(x,\tau_{1}) \right].
\end{equation}

\begin{figure}[t!!]
\centering
\includegraphics[width=\columnwidth]{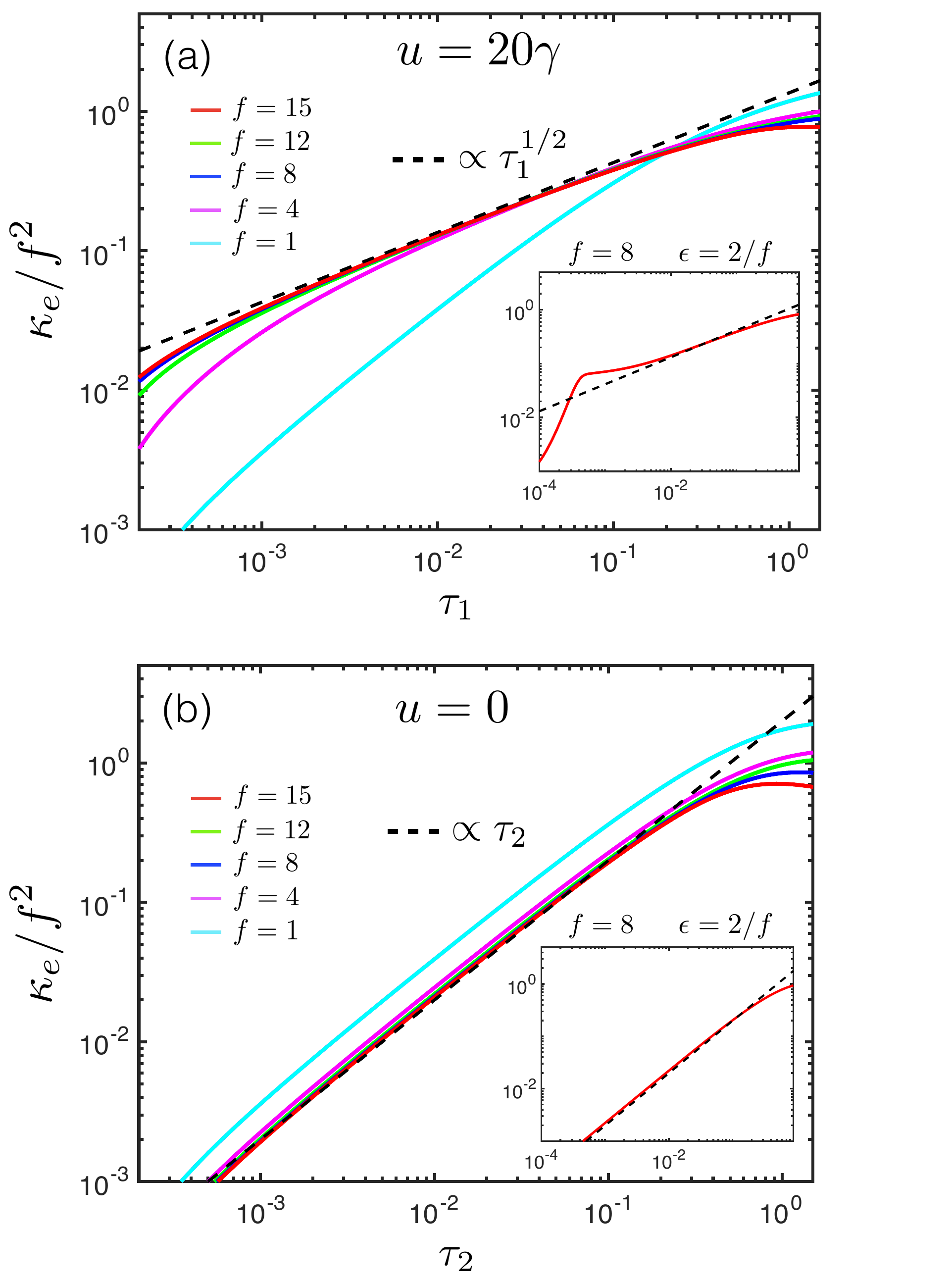}
\caption{\label{MFTKeTr} Dynamical evolution of the local particle fluctuations of the even sites for different fillings and $x_{e}=x_{o}=1$ in the two different regimes from \eqref{eq:er12} (solid color lines). The insets show the evolution for $x_{e}=1+\epsilon$ and $x_{o}=1-\epsilon$. Dashed black lines correspond to the analytical results \eqref{eq:er17} and \eqref{eq:er023}.}
\end{figure}

Ignoring the term of order $\gamma/uN$ in the denominator, Eq.~\eqref{eq:er15b} allows for a scaling solution of the form $p_{e}(x,\tau_{1})= g_{e}(\xi)/\tau_{1}^{\nu}$ with $\xi=x/\tau_{1}^{\nu}$, leading to 
\begin{equation}
\label{eq:er16}
p_{e(o)}(x,\tau_{1})= \frac{1}{4\Gamma(5/4) (x_{e} x_{o} \tau_{1})^{1/4}} e^{-\frac{(x-x_{e(o)})^{4}}{16\tau_{1} x_{e}x_{o}}} .
\end{equation}
This corresponds to anomalous diffusion of the probability distribution at short times. Using \eqref{eq:er16}, we can readily obtain the local number fluctuations $\kappa_{i}=\langle \hat{n}_{i}^{2} \rangle - \langle \hat{n}_{i} \rangle^{2}$, which reads
\begin{equation}
\label{eq:er17}
\frac{\kappa_{e(o)}}{f^{2}} =  \frac{\Gamma(3/4)}{\Gamma(5/4)} \sqrt{x_{e}x_{o}\tau_{1}}.
\end{equation}
This initial fast growth of the fluctuations can be understood as the system starting to explore neighboring configurations, separated by a small energy barrier from the initial state. Anomalous diffusion was also obtained in \cite{KollathPRL2013} for a BH model under the effects of local dephasing. This indicates that the impact of strong interactions on the dynamics is independent of their range of action and that the global nature of the dephasing in \eqref{eq:er8} does not play a major role within this level of approximation.

In Fig.~\ref{MFTKeTr}(a) we show the time evolution of the local particle number fluctuations of the even sites for $uN/\gamma \gg 1$ and $x_{e}=x_{o}=1$, resulting from numerical integration of \eqref{eq:er12}. We see that correlations do follow the power-law behavior predicted from the analysis in the continuous limit. As expected, the analytical results become more accurate for increasing values of the particle filling. In the inset, we show the evolution of the correlations for initial conditions with a finite imbalance $2\epsilon$ between even and odd sites, where $x_{e}=1+\epsilon$ and $x_{o}=1-\epsilon$. This imbalance results in a delayed approach of the algebraic regime \eqref{eq:er17}. This arises as the peaks of the effective diffusion distribution in \eqref{eq:er15b} and the initial probabilities are centered at different points. Thus, the probability distribution needs a certain amount of time to broaden before exploring the region of space that leads to \eqref{eq:er16}. For large enough $\epsilon$, the anomalous diffusive behavior \eqref{eq:er17} can get washed out if the broadening time required by $p_{e(o)}(x,\tau_{1})$ is larger than the time it takes to reach the reflective boundary at $x=0$, where $F_{e(o)}(x,\tau_{1}) \approx 0$ stops being a good approximation.

In the opposite limit, where interactions are of the same order, i.e.~$uN/\gamma \ll 1$, the behavior of the system becomes drastically different. Equation \eqref{eq:er12} becomes

\begin{align}
\label{eq:er19}
\partial_{\tau_{2}} p_{e(o)}(x,\tau_{2}) = \partial_{x} \Big[&\tilde{D}_{o(e)}(x,\tau_{2}) \partial_{x}p_{e(o)}(x,\tau_{2}) \nonumber \\ 
&- \tilde{F}_{o(e)}(x,\tau_{2}) p_{e(o)}(x,\tau_{2}) \Big] ,
\end{align}
with
\begin{align}
\label{eq:er20}
\tilde{D}_{i}(x,\tau_{2}) =& \int_{0}^{\infty} xy p_{i}(y,\tau_{2}) dy \nonumber \\
\tilde{F}_{i}(x,\tau_{2}) =& \int_{0}^{\infty} xy \partial_{y}p_{i}(y,\tau_{2}) dy \ ,
\end{align}
and $\tau_{2}=t/t_{2}^{*}$, where the new emergent time scale is $t_{2}^{*}= \frac{\gamma}{zJ^2}$. Assuming same initial conditions as before, for short times $\tau_{2} \ll 1$, the dynamics around the initial points reduces to
\begin{equation}
\label{eq:er21}
\partial_{\tau_{2}} p_{e(o)}(x,\tau_{2}) = x_{e} x_{o} \partial_{x}^{2} p_{e(o)}(x,\tau_{2}).
\end{equation}

Analogously to the previous case, this equation also allows for a scaling solution, leading to 

\begin{equation}
\label{eq:er022}
p_{e(o)}(x,\tau_{2})= \frac{1}{2\sqrt{\pi x_{e} x_{o} \tau_{2}}} e^{-\frac{(x-x_{e(o)})^2}{4x_{e} x_{o}\tau_{2}}}
\end{equation}
and local number fluctuations 

\begin{equation} 
\label{eq:er023}
\frac{\kappa_{e(o)}}{f^2}=2x_{e}x_{o}\tau_{2} .
\end{equation}

This corresponds to normal diffusion, characterized by the linear growth of $\kappa_{e}$ in time. The emergence of this regime in the limit $uN/\gamma \rightarrow 0$ can be understood in terms of the spectrum of the effective Hamiltonian \eqref{eq:er8b} for $J=0$. For $u=0$, all atomic states become degenerate. As a result, the system explores every configuration at the same rate, meaning that the evolution of the probability distribution is the same at every point in $x$, corresponding to a normal diffusion process in configuration space. In Fig.~\ref{MFTKeTr}(b) we present the evolution of $\kappa_{e}/f^{2}$ for $uN/\gamma =0$. We find good agreement between numerical results and \eqref{eq:er023} that improves for larger filling $f$. We also see that an initial imbalance does not modify the evolution of the correlations, which follows from the diffusion function being homogeneous, i.e.~it has the same form independently of the initial point of the neighbouring site.

\begin{figure}[t!!]
\centering
\includegraphics[width=\columnwidth]{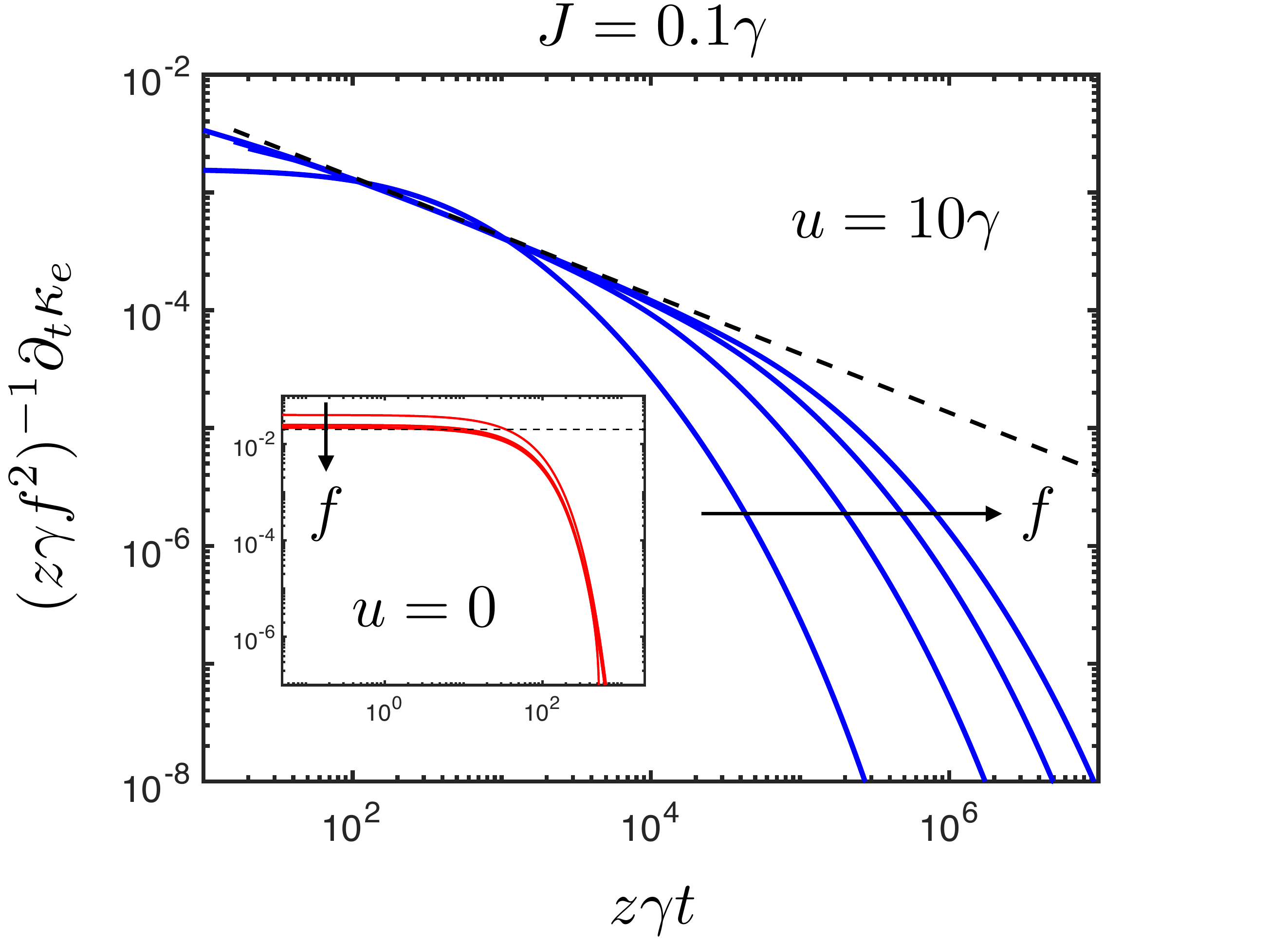}
\caption{\label{PlotdK}Dynamical evolution of the time derivative of the local particle number fluctuations for different values of the filling $f=1,4,8,12$. We observe how for $uN/\gamma \gg 1$ the time-scales at which the steady state is reached become larger for increasing filling, consequence of the large energy splitting with the states associated with high on-site population ($x \gg 1$). In contrast, for $uf/\gamma \ll 1$, the steady state is approached orders of magnitude faster as all configurations are energetically degenerate. Dashed black lines correspond to the analytical predictions for short times \eqref{eq:er17} and \eqref{eq:er023}.}
\end{figure}

The long time behavior of the system can also be understood in terms of the energy spectrum of \eqref{eq:er8b}. For $uN/\gamma \gg 1$, after the distribution reaches the $x=0$ boundary, the dynamics becomes dominated by the $x \gg 1$ region. These configurations are associated to states with large occupation numbers and have a high energy cost to populate. As a result, there is a slowdown of the dynamics, which becomes more pronounced for higher fillings. This is shown in Fig.~\ref{PlotdK}, where one can see how for higher values of $f$ the approach of the steady state becomes increasingly slower. This phenomenon was also observed in \cite{KollathPRL2013}, where it is shown that correlations exhibit a stretched exponential behavior $\kappa_{e}(\tau_{1}) \propto e^{-\alpha\sqrt{\tau_{1}}}$. For $uN/\gamma \ll 1$, this is no longer the case due to the energy degeneracy among the different atomic configurations, which allows the probability distribution to explore all the configuration space at the same rate, leading to a much faster approach of the steady state distribution. This behavior can be observed in the inset of Fig.~\ref{PlotdK}, where the time at which the steady state is reached is orders of magnitude smaller than in the $uN/\gamma \gg 1$ case.

\subsection{Large hopping limit $J \gg (U,U_{l},\gamma)$}
\label{BCRlJ}

We now return to Eq.~\eqref{eq:er8} and consider the limit where the hopping amplitude is the dominant energy scale $J \gg (U,U_{l},\gamma)$. For convenience, we will consider vanishing onsite interactions $U=0$. In this regime, it is best to work in momentum space, so we start with Eq.~\eqref{eq:er4} in the limit $J \ll (|\Delta|,\kappa)$. To simplify the problem, we move to an interaction picture, taking $\hat{H}_{J}$ as the free Hamiltonian, and perform a RWA, i.e.~we eliminate all the left over rotating terms. 
This yields
\begin{align}
\label{eq:er22}
\dot{\hat{\rho}}_{\textrm{at}} =& (-i)[\hat{H}_{\textrm{eff}}, \hat{\rho}_{\textrm{at}}] \nonumber \\
& + \gamma \sum_{k=0}^{k<\pi} \Big\{ \mathcal{D}[\hat{b}^{\dagger}_{k}\hat{b}_{k-k_{\pi}}] (\hat{\rho}_{\textrm{at}}) + \mathcal{D}[\hat{b}^{\dagger}_{k-k_{\pi}}\hat{b}_{k}] (\hat{\rho}_{\textrm{at}}) \Big\}, \\
\hat{H}_{\textrm{eff}} = & \sum_{k=0}^{k<\pi} \Big[  (\varepsilon_{k} + \delta) \hat{n}_{k}  + (\varepsilon_{k-k_{\pi}} + \delta) \hat{n}_{k-k_{\pi}} \nonumber \\ 
 &+2\delta \hat{n}_{k} \hat{n}_{k-k_{\pi}} \Big] , 
\end{align}
with $\hat{n}_{k}=\hat{b}_{k}^{\dagger}\hat{b}_{k}$ and $\delta=g^2\Delta/(\Delta^{2}+ \kappa^{2}/4)$. 

From the definition of $\hat{\Phi}$ we know that this operator only couples pairs of momentum states with momentum differing by $k_{\pi}$. For $U=0$, this means that $[\hat{n}_{k}+\hat{n}_{k-k_{\pi}},\hat{H}_{\textrm{eff}}]=0$. Together with $[\hat{n}_{k}+\hat{n}_{k-k_{\pi}},\hat{b}^{\dagger}_{k}\hat{b}_{k-k_{\pi}}]=[\hat{n}_{k}+\hat{n}_{k-k_{\pi}},\hat{b}^{\dagger}_{k-k_{\pi}}\hat{b}_{k}]=0$, we find that the dynamics of each pair of momentum states is decoupled from that of the other momentum modes. For simplicity, we focus only on the states $k_{0}=0$ and $k_{\pi}=\pi$, with $\hat{H}_{J}=-J(\hat{n}_{k_{0}} - \hat{n}_{k_{\pi}})$ and $\hat{\Phi}= \hat{b}^{\dagger}_{k_{0}} \hat{b}_{k_{\pi}} + \hat{b}^{\dagger}_{k_{\pi}} \hat{b}_{k_{0}}$. The reduced version of \eqref{eq:er22} reads
\begin{equation}
\label{eq:er23}
\dot{\hat{\rho}}_{\textrm{at}} = (-i)[\hat{H}_{\textrm{eff}},\hat{\rho}_{\textrm{at}}] + \gamma \left\{ \mathcal{D}[\hat{b}^{\dagger}_{k_{0}}\hat{b}_{k_{\pi}}] (\hat{\rho}_{\textrm{at}}) + \mathcal{D}[\hat{b}^{\dagger}_{k_{\pi}} \hat{b}_{k_{0}}] (\hat{\rho}_{\textrm{at}}) \right\}
\end{equation}
with
\begin{equation}
\label{eq:er24}
\hat{H}_{\textrm{eff}} = N\delta  - J(\hat{n}_{k_{0}} - \hat{n}_{k_{\pi}}) + 2\delta \hat{n}_{k_{0}} \hat{n}_{k_{\pi}} . 
\end{equation}
As result of the RWA, the long-ranged interactions have been reduced to a number-number interaction between the momentum states $k_{0}$ and $k_{\pi}$. This interaction modifies the spectrum of $\hat{H}_{\textrm{eff}}$ but commutes with the rest of the terms in \eqref{eq:er24}. The form of \eqref{eq:er24} leads to momentum being a good quantum number $[\hat{n}_{k},\hat{H}_{\textrm{eff}}]=0$. This feature enables us to recast Eq.~\eqref{eq:er23} as a linear rate equation of the form $\dot{P}_{i} = \sum_{j} (\Gamma_{j \rightarrow i} P_{j} - \Gamma_{ i \rightarrow j} P_{i})$, where $P_{i}$ is the probability for the system to be in a certain state $|n_{k_{0}},n_{k_{\pi}}\rangle$ and $\Gamma_{j \rightarrow i}$ are the transition rates between different states. This equation of motion depends linearly on the state probabilities, as opposed to the non-linear form of \eqref{eq:er12}. The general solution for such a rate equation corresponds to a linear combination of exponential decays. Thus, in contrast to the results of the previous section, the relaxation towards the steady state is now exponential. Note that $\Gamma_{j \rightarrow i}=\gamma$ for all $i$,$j$, leading to $P_{i}(t \rightarrow \infty)= 1/D$, being $D=N+1$ the size of the Hilbert space, for all $i$, i.e.~the steady state is indeed an effective infinite temperature state as expected. In terms of average population, this corresponds to the atoms equally occupying each momentum mode $\langle \hat{n}_{k_{0}} \rangle = \langle \hat{n}_{k_{\pi}} \rangle=N/2$ in the steady state.

\begin{figure}[t]
\includegraphics[width=\columnwidth]{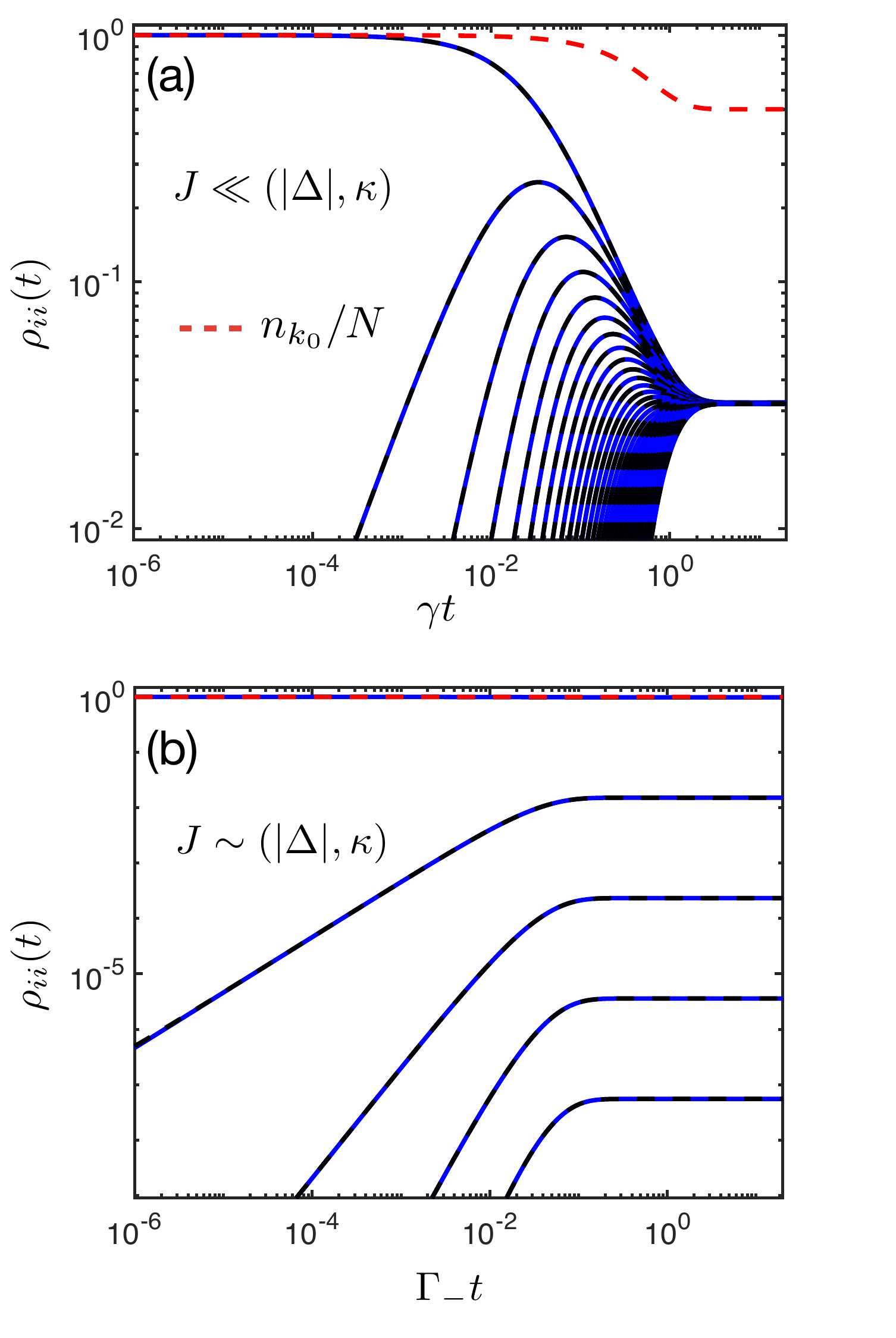}
\caption{\label{DynamicsRWA} Dynamical evolution of the diagonal density matrix elements for $N=30$ and $U=0$, in the bad (a) and good (b) cavity regimes. The blue lines denote the solution obtained from \eqref{eq:er23} and the dashed black lines from \eqref{eq:er1b}. In red, the time evolution of the average number of particles in the $k_{0}$ state. Parameters: (a) $g=10^{-4}\kappa$, $\Delta=-2\kappa$, $J=10^{-4}\kappa$; (b) $g=10^{-4} \kappa$, $\Delta=-2\kappa$, $J=\kappa$. In both cases all particles are initially  in the momentum state $k_{0}=0$.}
\end{figure}

In Fig.~\ref{DynamicsRWA}(a) we present the evolution of the diagonal density matrix elements from the numerical integration of \eqref{eq:er23} (solid blue) and the full quantum master equation \eqref{eq:er1b} (dashed black). As expected, all the diagonal elements exponentially approach the same steady state value and the population of the $k_{0}$ mode becomes $N/2$. The match between both numerical solutions indicate the validity of the RWA for large hopping $J$.

We have checked numerically that small corrections due to the rotating terms will only account for small oscillations at very short times. The effect of small but finite $U$ is that of slowly spreading the atoms across all the available momentum states. We expect that as the atoms start populating other pairs of momentum modes differing by $k_{\pi}$, these will quickly thermalize into the steady state discussed above and eventually scatter towards other states at a rate of order $U^{2}/J \ll \gamma$. Once the atoms have explored all possible momentum modes, the final state will correspond to an infinite temperature state that spans the entire Hilbert space of the system, corresponding to the steady state of Eq.~\eqref{eq:er8}.

\section{Good cavity regime}
\label{GCR}

This regime is characterized by the cavity being able to resolve the scales of atomic transitions, i.e. $J\sim(|\Delta|,\kappa)$. Thus, our starting point is Eq.~\eqref{eq:er4}. In this limit, it is only possible to bring this equation into Lindblad form by applying the RWA. For our purposes, we will again consider the limit of vanishing on-site interactions $U=0$, in order to focus on the impact of the cavity. This leads to
\begin{align}
\label{eq:er25}
\dot{\hat{\rho}}_{\textrm{at}} =& (-i)[\hat{H}_{\textrm{eff}},\hat{\rho}_{\textrm{at}}] \nonumber \\
&+ \sum_{k=0}^{k<\pi} \left\{ \Gamma_{k,+} \mathcal{D}[\hat{b}^{\dagger}_{k}\hat{b}_{k-k_{\pi}}] (\hat{\rho}_{\textrm{at}}) +  \Gamma_{k,-} \mathcal{D}[\hat{b}^{\dagger}_{k-k_{\pi}} \hat{b}_{k}] (\hat{\rho}_{\textrm{at}}) \right\} ,
\end{align}
with
\begin{align}
\label{eq:er25b}
 \hat{H}_{\textrm{eff}} = & \sum_{k=0}^{k<\pi} \Big[  (\varepsilon_{k} + \delta_{k,-}) \hat{n}_{k}  + (\varepsilon_{k-k_{\pi}} + \delta_{k,+}) \hat{n}_{k-k_{\pi}} \nonumber \\ 
 &+\lambda_{k} \hat{n}_{k} \hat{n}_{k-k_{\pi}} \Big], 
\end{align} 
where $\delta_{k,\pm} = g^2 \textrm{Im}[G(\mp\varepsilon_{k}\pm\varepsilon_{k-k_{\pi}})]$, $\lambda_{k}= \delta_{k,+}+\delta_{k,-}$ and $\Gamma_{k,\pm}=2  g^2\textrm{Re}[G(\mp\varepsilon_{k}\pm\varepsilon_{k-k_{\pi}})]$.

We then reduce our description to a pair of representative momentum states $(k_{0},k_{\pi})$, which reads
\begin{equation}
\label{eq:er26}
\dot{\hat{\rho}}_{\textrm{at}} = (-i)[\hat{H}_{\textrm{eff}},\hat{\rho}_{\textrm{at}}] + \Gamma_{-} \mathcal{D}[\hat{b}^{\dagger}_{k_{0}}\hat{b}_{k_{\pi}}] (\hat{\rho}_{\textrm{at}}) +  \Gamma_{+} \mathcal{D}[\hat{b}^{\dagger}_{k_{\pi}} \hat{b}_{k_{0}}] (\hat{\rho}_{\textrm{at}}) 
\end{equation}
with
\begin{equation}
\label{eq:er26b}
\hat{H}_{\textrm{eff}} = (\delta_{-} - J)\hat{n}_{k_{0}} + (\delta_{+}+J)\hat{n}_{k_{\pi}} + \lambda \hat{n}_{k_{0}} \hat{n}_{k_{\pi}} , 
\end{equation}
where $\delta_{\pm}=  \frac{g^2(\Delta \pm 2J)}{(\Delta \pm 2J)^2 + \kappa^2/4}$, $\lambda = \delta_{+} + \delta_{-}$ and $\Gamma_{\pm}= \frac{g^2 \kappa}{(\Delta \mp 2J)^2 + \kappa^2/4)}$ are the transition rates. Analogously to \eqref{eq:er24}, here momentum is also a good quantum number, and thus the dynamics of the states $|n_{k_{0}},n_{k_{\pi}}\rangle$ will be purely dissipative. This again allows to rewrite the master equation \eqref{eq:er26} as a rate equation, where now the transitions $|n_{k_{0}},n_{k_{\pi}}\rangle \rightarrow |n_{k_{0}}+1,n_{k_{\pi}}-1\rangle$ occur at rate $\Gamma_{-}$ and the transitions $|n_{k_{0}},n_{k_{\pi}}\rangle \rightarrow |n_{k_{0}}-1,n_{k_{\pi}}+1\rangle$ at rate $\Gamma_{+}$, leading to the rates $\Gamma_{i \rightarrow j}$ satisfying detailed balance
\begin{equation}
\label{eq:er27}
\frac{\Gamma_{-}}{\Gamma_{+}}=\frac{\frac{\kappa^2}{4} + (2J - \Delta)^2}{\frac{\kappa^2}{4} + (2J + \Delta)^2}=e^{\frac{2J}{T_{\textrm{eff}}}} \ .
\end{equation}

\begin{figure}[t]
\includegraphics[width=\columnwidth]{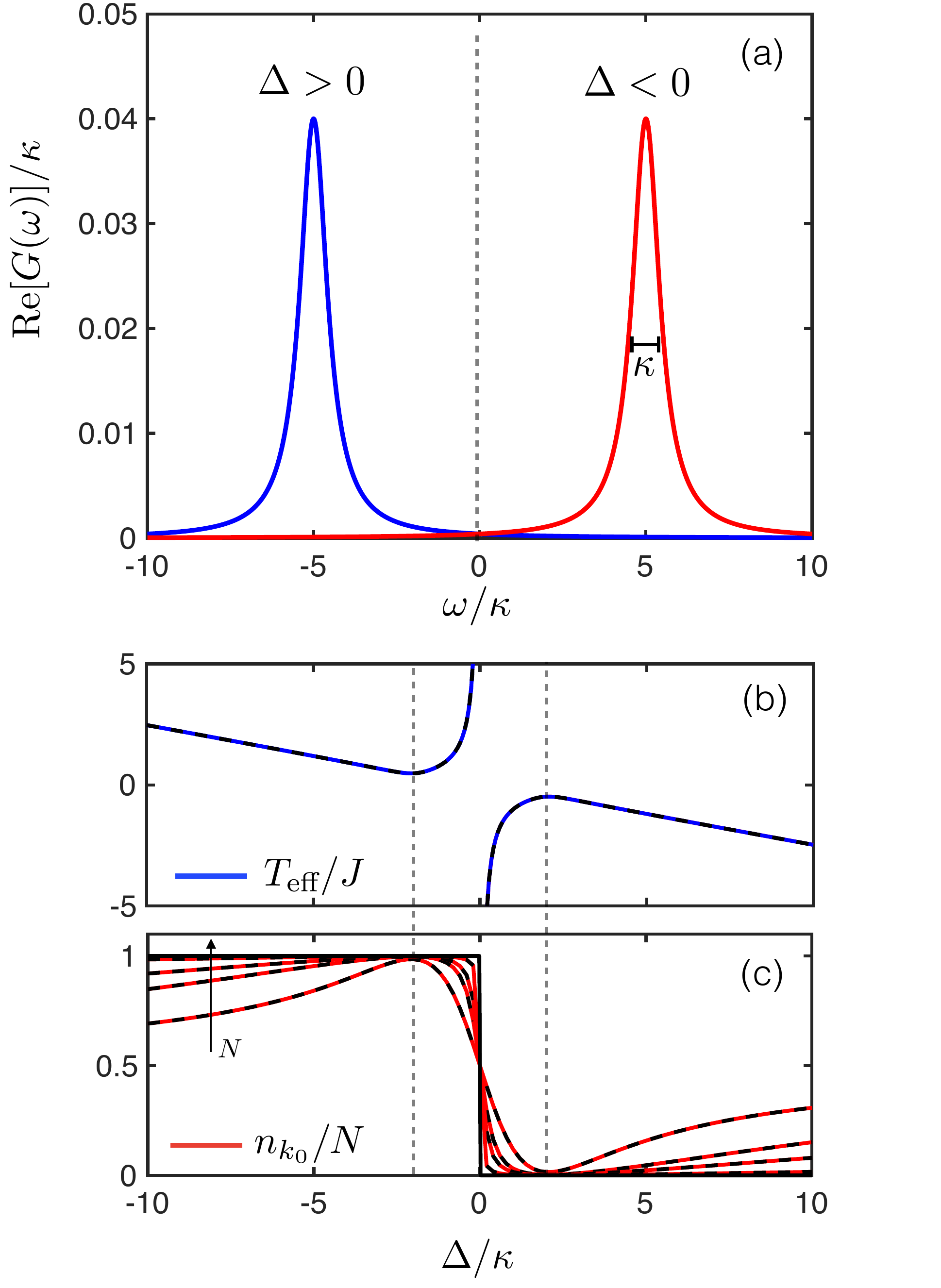}
\caption{\label{Figure3}(a) The spectral function $\textrm{Re}[G(\omega)]$ of the cavity field. For zero-temperature white noise, this corresponds to a Lorentzian function of width $\kappa$ centered at $-\Delta$. For positive(negative) values of $\Delta$, i.e.~a blue(red) detuned light field, we obtain pumping of the high(low) momentum state. (b) The effective steady state temperature $T_{\textrm{eff}}$ as a function of $\Delta/\kappa$ with $J=\kappa$. (c) Average number of particles in the low momentum state $\langle \hat{n}_{k_{0}} \rangle /N$ as a function of $\Delta/\kappa$, with $J=\kappa$, for $N=1,5,10,50$ (dashed) and $N \rightarrow \infty$ (solid). The function becomes sharper for increasing $N$ and develops a step at $\Delta=0$, where it always satisfies $\langle \hat{n}_{k_{0}} \rangle/N = 1/2$. Vertical grey dashed lines indicate $\Delta=\pm 2J$.}
\end{figure}

Note that the form of these rates follow the Lorentzian form of cavity spectral function (see Fig.~\ref{Figure3}(a)). In \eqref{eq:er27}, $T_{\textrm{eff}}$ stands for the effective temperature of the system. The steady state of the system is then given by a thermal distribution whose temperature can be controlled via the detuning $\Delta$. This enables control over the population of the atoms which can be optically pumped into either momentum mode according to 

\begin{equation}
\label{eq:er28} \langle \hat{n}_{k_{0}} \rangle=\frac{r-(N+1)r^{N+1}+Nr^{N+2}}{(1-r)(1-r^{N+1})},
\end{equation}
for $r \neq 1$, with $r=\Gamma_{-}/\Gamma_{+}$, and $\langle \hat{n}_{k_{\pi}} \rangle=N-\langle \hat{n}_{k_{0}} \rangle$. The case $r=1$ corresponds to the bad cavity limit result ($\Gamma_{+}=\Gamma_{-}=\gamma$) with $T_{\textrm{eff}}=\infty$ and $\langle \hat{n}_{k_{0}} \rangle =\frac{N}{2}$. In Fig.~\ref{Figure3}(b) and (c), we present the effective temperature and the population fraction of the $k_{0}$ mode as a function of the detuning. The dashed lines correspond to the results in Eqs.~\eqref{eq:er27} and \eqref{eq:er28}, and the solid lines to the results from numerical integration of \eqref{eq:er1b}. Note that as the number of particles $N$ becomes very large the behavior of $\langle \hat{n}_{k_{0}} \rangle$ becomes extreme, i.e.~the atoms fully polarize in either state only as a function of the sign of $\Delta$. This follows from considering $N \rightarrow \infty$ in \eqref{eq:er28} which yields $\langle \hat{n}_{k_{0}} \rangle/N \rightarrow 1$ for $\Delta<0$, $\langle \hat{n}_{k_{0}} \rangle/N \rightarrow 0$ for $\Delta>0$, and $\langle \hat{n}_{k_{0}} \rangle/N \rightarrow 1/2$ when $\Delta=0$ or in the bad cavity limit $|\Delta|/J \gg 1$.

The dynamical behavior of \eqref{eq:er26} is shown in Fig.~\ref{DynamicsRWA}(b). The system is initialized with all atoms in the $k_{0}$ state and with $\Delta=-2J$. As expected, the atoms mostly remain in this configuration when they approach the steady state. This can also be observed in the behavior of $\langle \hat{n}_{k_{0}} \rangle$ which remains very close to 1 throughout the entire evolution. As previously, the good match with the solutions obtained from \eqref{eq:er1b} indicates the validity of our approximations.

The presence of optical pumping in this regime is interesting from the perspective of state preparation, as it could be used as a cavity cooling mechanism to maximize the number of particles in the BEC state. A similar strategy was used in \cite{HemmerichNat2012} for cavity cooling by pumping the system along the cavity axis. The main difference with \cite{HemmerichNat2012} is that here the energy splitting between atomic transitions is not given by the recoil energy of the atoms but instead by the hopping amplitude $J$. Analogously to Sec.~\ref{BCRlJ}, we expect the effects of finite but small $U$ to be that of slowly scattering the atoms into other momentum configurations at a rate $U^{2}/J \ll \Gamma_{\pm}$ and a final state corresponding to each pair of momentum modes being equally populated but obeying a thermal distribution between the two modes.

\section{Conclusions}
\label{Conc}

In this paper, we have studied the non-equilibrium dynamics of a gas of ultracold atoms inside an optical resonator, in the presence of an optical lattice and transverse driving. We have shown that the relaxation dynamics of ultracold atomic inside optical cavities can display a wide variety of different behaviors, as a consequence of the competition between interactions, hopping and dissipation.

In the bad cavity regime, we obtained that the steady state always corresponds to infinite temperature. However, the approach to this steady state strongly depends on the considered parameter regime. For small hopping, we integrated the coherences of the density matrix to obtain a description only in terms of the probabilities associated to each atomic configuration and analyzed these quantities using a Gutzwiller ansatz and considering a continuum description in the limit of large filling $f$. The result was an algebraic decay of the particle number fluctuations, which can be associated with anomalous diffusion if one of the interaction strengths dominates over the other, or with normal diffusion when the interactions are of the same order. For large hopping, by performing a RWA, we obtained that the dynamics of the system is entirely dissipative and shown that the resulting master equation can be mapped into a linear rate equation. As a consequence, in this limit the approach of the steady state is given by a linear combination of exponential decays.

In contrast, in the good cavity regime, we found that for vanishing on-site interactions, the system evolves into a different steady state, given by a thermal distribution between pairs of momentum states. This allows for optical pumping between these pairs as the effective temperature can be controlled using the detuning $\Delta$. This could be implemented as an alternative scheme for cavity-assisted cooling of atomic clouds in ultra-narrow band cavities, where one can access good cavity regime.

Finally, we believe that all the presented results are within experimental reach. Observation of the algebraic regime requires $(U,U_{l},\gamma) \gg J$ as realized in \cite{Esslinger2}, with $(U,U_{l})=10J$-$40J$. This can be tailored to $uN \ll \gamma$ or $uN \gg \gamma$, while satisfying the single-band approximation \cite{Esslinger2}, i.e.~all energy scales being much smaller that the interband energy gap. Tuning of the dissipation rate is available through $\gamma \propto V_{2D}/\Delta^2$, where $V_{2D}$ is the optical lattice depth, meaning that the ratio $\gamma/J$ can be controlled. The regime of exponential relaxation could be observed by reducing the effective short-range interactions by means of Feshbach resonances, using external magnetic fields. Implementation of the bad cavity regime requires detuning the pump laser such that $|\Delta| \gg J$, and using an optical cavity with a linewidth of order $\kappa \sim$ MHz. Both conditions are already fulfilled in \cite{Esslinger2}, where $\Delta \sim$ MHz and $J \sim$ Hz.  Exploring the good cavity regime requires a detuning on the order of $\Delta \sim$ Hz and an ultranarrow-band optical cavity \cite{HemmerichPRL2014,HemmerichPNAS2014,HemmerichNat2012}, where $\kappa \sim$ Hz.

\section{Acknowledgements}
\label{Ack}

We are grateful to Nigel Cooper, Austen Lamacraft, Christopher Parmee and Ulrich Schneider for fruitful and stimulating discussions.
E.I.R.C.~acknowledges support from the Winton Programme for the Physics of Sustainability and the UK Engineering and Physical Sciences Research Council (EPSRC).
A.N.~holds a University Research Fellowship from the Royal Society and acknowledges additional support from the Winton Programme for the Physics of Sustainability.

\bibliography{library}

\end{document}